\documentstyle[preprint,eqsecnum,aps,epsfig]{revtex}
\tightenlines

 \def\CJ{{\cal J}}  
  \def\CO{{\cal O}} 
   \def\CT{{\cal T}}
   
   \def\la{\langle}
\def\ra{\rangle}
\def\bra#1{\left\la #1\right|}
\def\ket#1{\left| #1\right\ra}

\def\abs#1{\left| #1\right|}

\def\({\left(}
\def\){\right)}
\def\[{\left[}
\def\]{\right]}

\def\Re{{\rm Re \,}}
\def\Im{{\rm Im \,}}

\def\half{\frac{1}{2}}
\def\bhalf{{1\over 2}}
\def\del{\partial}

\def\slash#1{\rlap{$#1$}/} 
\def\fpi{f_\pi}

\def\mpi{m_\pi}
\def\vect#1{\mbox{\boldmath$#1$}}

\def\ben{\begin{equation}}
\def\een{\end{equation}}
\def\bey{\begin{eqnarray}}
\def\eey{\end{eqnarray}}
\def\beyn{\begin{eqnarray*}}
\def\eeyn{\end{eqnarray*}}
\def\ba{\begin{array}}
\def\ea{\end{array}}
\def\bleft{\begin{flushleft}}
\def\eleft{\end{flushleft}}
\def\bright{\begin{flushright}}
\def\eright{\end{flushright}}
\def\bcenter{\begin{center}}
\def\ecenter{\end{center}}

\def\etal{{\it et al.}}

\begin{document}
\draft
\preprint{
  \begin{tabular}{r}
    DPNU-99-02\\
    January 1999
  \end{tabular}
}

\title{Spin-Polarizabilities of the Nucleon in Chiral Soliton Model
with Dispersion Relation}

\author{Y. Tanushi\thanks{E-mail: tanushi@nuc-th.phys.nagoya-u.ac.jp}
    and S. Saito\thanks{E-mail: saito@nuc-th.phys.nagoya-u.ac.jp}}
\address{Department of Physics, Nagoya University,
Nagoya 464-8602, Japan}
\author{M. Uehara\thanks{E-mail: ueharam@cc.saga-u.ac.jp}}
\address{Department of Physics, Saga University,
Saga 840-8502, Japan}
\date{\today}
\maketitle

\begin{abstract}
We calculate the spin-polarizabilities of the nucleon
by means of dispersion relation, where the pion photoproduction
amplitude predicted by chiral soliton model is
utilized.
We consider the $N\pi$ and $\Delta\pi$ channels
in the pion-photoproduction amplitude, and also
evaluate the contribution of the anomalous term of the
$\pi^0\gamma\gamma$ process.
In a narrow decay-width limit of the $\Delta$ particle
the result coincides with that in heavy baryon chiral
perturbation theory (HBChPT) except for the interference part of
the electric and magnetic amplitudes.
A comparison with the multipole analysis is given.
\end{abstract}


\section{Introduction}
\label{sec:sec1}

The electromagnetic polarizabilities are the quantities to represent
the response of the nucleon to external electromagnetic filed
and reflect its internal structure.
The electric and magnetic polarizabilities $\alpha$ and $\beta$
are measured by elastic Compton scattering with unpolarized photon and
nucleon,
and many experiments have been already devoted to the study.

Recently experiments with polarized photon beam
and proton target have been available,
and it is expected that much information about the spin-dependent structure
can be obtained from low energy polarized Compton scattering.
The celebrated Drell-Hearn-Gerasimov sum rule \cite{DHG}
is given by the spin-dependent photoabsorption cross section
$\sigma_{abs}^\lambda$ with the total helicity $\lambda$ as follows:
\ben
-{e^2\kappa^2\over 8\pi M^2}
={1\over 4\pi^2}\int_{\omega_{\,\mathrm{th}}}^\infty{{\rm
d}\omega\over\omega}\,
\(\sigma_{abs}^{1/2}(\omega)-\sigma_{abs}^{3/2}(\omega)\),
\een
where $\omega_{\,\mathrm{th}}$ is the threshold energy, $\kappa$
the anomalous magnetic moment of the nucleon, and $M$ the nucleon mass.
The forward spin-polarizability $\gamma_0$ is given by
the once-subtracting sum rule,
the Gell-Mann-Goldberger-Thirring sum rule\cite{GGT},
with the inverse weight of $\omega^3$ as follows:
\ben
\gamma_0
={1\over 4\pi^2}\int_{\omega_{\,\mathrm{th}}}^\infty{{\rm
d}\omega\over\omega^3}\,
\(\sigma_{abs}^{1/2}(\omega)-\sigma_{abs}^{3/2}(\omega)\).
\label{eq:GGTSR}
\een
Recently, Ragusa showed \cite{Ragusa}  that
there are four independent spin-polarizabilities $\gamma_i$($i=1,\ldots, 4$)
in the amplitude of $\CO(\omega^3)$,
and that the forward spin-polarizability $\gamma_0$ is given by
$\gamma_0=\gamma_1-\gamma_2-2\gamma_4$.
There has not yet been  any experiment about the spin-polarizabilities.

Theoretical investigations on the spin-polarizabilities have been
carried out within the framework of
heavy-baryon chiral perturbation theory (HBChPT):
The leading order terms from the $N\pi$ loops
are of $1/m_\pi^2$, which show the importance
of the pion cloud around the nucleon for the polarizabilities.
However, the result contradicts with and has an opposite sign
to that of the pion photoproduction multipole
analysis\cite{Sandorfi94}.
Bernard \etal\cite{Bernard92a} showed that the contribution of
the $\Delta$-pole terms
is negative and reduces largely that of the $N\pi$-loop terms:
$\gamma_0^{N\pi-loop}+\gamma_0^{\Delta-pole}
=(4.44-3.66)\times10^{-4}{\rm fm^4}$.
The sum is $0.78\times10^{-4}{\rm fm^4}$, while the multipole analysis
yields $\gamma_0=-1.34(-0.38)\times10^{-4}{\rm fm^4}$
for the proton (neutron)\cite{Sandorfi94}.
Similar results are obtained by Hemmert \etal\cite{Hemmert97,Hemmert98} ,
within a small scale expansion framework, where the small scale
is either of the pion mass,
nonrelativistic momentum or the mass difference between the $\Delta$
particle and the nucleon, and the $\Delta$ particle is treated as  explicit
degrees of freedom.
In the work the spin-polarizabilities $\gamma_1,\ldots,\gamma_4$
are calculated in detail and compared with the result of the
multipole analysis.

In a previous paper\cite{Tanushi} we have calculated
the forward spin-polarizability $\gamma_0$
with use of dispersion relation in the chiral soliton model.
This followed the calculation of the electric and magnetic
polarizabilities in the same context\cite{Saito95}.
Chiral soliton model is a QCD motivated one based on the idea
of large $N_c$
and of the spontaneous breaking of chiral symmetry\cite{Adkins}.
The electromagnetic
polarizabilities are  sensitive to the pion cloud around the
nucleon, so that the model may be well-suited to the study of the
polarizabilities. The pion photoproduction Born amplitudes
obtained by the model have been
shown to satisfy  the low-energy theorem \cite{Saito93}, and were employed
for calculating the dispersion integrals.
Further, we note that the $\Delta(1232)$ state is a rotational excited
state of the soliton, so that they are naturally treated as
an equal partner with each other.
In our approach, the chiral soliton
model is understood as an effective model to derive the pion-photoproduction
amplitude.
The imaginary part of the amplitude of  Compton scattering
can be derived with use of the unitarity condition, and the polarizabilities
are then calculated with dispersion relation, which is
considered to be a means to calculate loop integrals
\cite{Bernard92b,Kniehl}.

In Ref.\cite{Tanushi}
we found that the contributions from the electric and magnetic parts
in the $N\pi$ channel coincide with those from
the $N\pi$-loop and $\Delta$-pole terms  calculated
in HBChPT, if we take the narrow width limit of the $\Delta$ state.
In this paper we study the spin-polarizabilities $\gamma_1,
\ldots, \gamma_4$.
It is shown that the imaginary parts of the structure functions of
Compton scattering amplitude, $A_5(\omega,\theta=0)$ and
$A_6(\omega,\theta=0)$, turn out to be zero.
Naive application of the dispersion relation results in that
$\gamma_3$ and $\gamma_4$ vanish.
We argue that the convergent dispersion integral does not necessarily
mean no need of the subtraction, and discuss how to remedy the ill-defined
dispersion integrals.

In section \ref{sec:sec2} we give the dispersion relations for
the polarizabilities.
For  this paper to be self-contained
the results on the spin-independent polarizabilities
are summarized. Section \ref{sec:sec3} is devoted to the calculation
of the imaginary parts of Compton scattering amplitude with use
of the pion photoproduction amplitude to the $N\pi$ and $\Delta\pi$
channels.
The $\pi_0\gamma\gamma$ anomalous contribution is given in section
\ref{sec:sec4}.
Numerical results and discussion are given in section \ref{sec:sec5}.

\section{Dispersion Relations of the Amplitudes}
\label{sec:sec2}

The electromagnetic polarizabilities of the nucleon are defined by
Compton scattering amplitude, which is represented, at the
center-of-mass system, as\cite{Hemmert98}
\bey
f&=&
 A_1(\omega,\theta)
   \vect\epsilon^{\prime *}\!\cdot\!\vect\epsilon
+A_2(\omega,\theta)
   (\vect\epsilon^{\prime *}\!\cdot\!\hat{\vect k})
   (\vect\epsilon\!\cdot\!\hat{\vect k}') \nonumber \\
&&+A_3(\omega,\theta)\, i\vect\sigma\!\cdot\!
     (\vect\epsilon^{\prime *}\!\times\!\vect\epsilon)
  +A_4(\omega,\theta)\, i\vect\sigma\!\cdot\!
     (\hat{\vect k}'\!\times\!\hat{\vect k})
     (\vect\epsilon^{\prime *}\!\cdot\!\vect\epsilon) \nonumber \\
&&+A_5(\omega,\theta)\, i\vect\sigma\!\cdot\!
     \[(\vect\epsilon^{\prime *}\!\times\!\hat{\vect k})
      (\vect\epsilon\!\cdot\!\hat{\vect k}')
     -(\vect\epsilon\!\times\!\hat{\vect k}')
      (\vect\epsilon^{\prime *}\!\cdot\!\hat{\vect k})\] \nonumber \\
&&+A_6(\omega,\theta)\, i\vect\sigma\!\cdot\!
    \[(\vect\epsilon^{\prime *}\!\times\!\hat{\vect k}')
      (\vect\epsilon\!\cdot\!\hat{\vect k}')
     -(\vect\epsilon\!\times\!\hat{\vect k})
      (\vect\epsilon^{\prime *}\!\cdot\!\hat{\vect k})\],
  \label{eq:genCom}
\eey
where $\vect\epsilon (\vect\epsilon ')$ and
$\hat{\vect k}(\hat{\vect k'})$ are the polarization
vector and the unit momentum of the incident(outgoing) photon, respectively.
$\vect\sigma$ denotes the Pauli matrix of the nucleon,
so that the first and second terms are spin-independent,
while the others are spin-dependent.
The independent structure functions $A_i(\omega,\theta)$
with $i=1,\ldots,6$ are functions of
the photon energy $\omega(=\omega')$ and the scattering angle $\theta$.

The above structure functions can be divided into the Born,  non-Born and
anomalous terms, respectively, as follows:
\ben
A_i(\omega,\theta)
=A_i^B(\omega,\theta)+A_i^{nB}(\omega,\theta)
+A_i^{{\mathrm{anom}}}(\omega,\theta)
\een
with $i=1,\ldots,6$. The anomalous term is given by the $\pi_0\gamma\gamma$
process through the Wess-Zumino-Witten term\cite{Wess}.
The Born terms represent the scattering from a spin 1/2 point particle
and are determined by its mass, charge,
and anomalous magnetic moment,
while the non-Born terms reflect its internal structure.
The polarizabilities appear in the low-energy expansion of
the non-Born and  anomalous terms of the amplitudes.
The electric and magnetic polarizabilities $\alpha$ and $\beta$
are defined as the coefficients of the terms of $\CO(\omega^2)$,
while the spin-polarizabilities $\gamma_i$ are those of $\CO(\omega^3)$:
\bey
A_1(\omega,\theta)&=& A_1^B(\omega,\theta)+
(\alpha+\beta\cos\theta)\, \omega^2+\CO(\omega^3), \nonumber \\
A_2(\omega,\theta)&=& A_2^B(\omega,\theta)
-\beta\,\omega^2+\CO(\omega^3), \nonumber \\
A_3(\omega,\theta)&=& A_3^B(\omega,\theta)+
[\gamma_1-(\gamma_2+2\gamma_4)\cos\theta]\,\omega^3+\CO(\omega^4), \nonumber \\
A_4(\omega,\theta)&=& A_4^B(\omega,\theta)+
\gamma_2\,\omega^3+\CO(\omega^4), \nonumber \\
A_5(\omega,\theta)&=& A_5^B(\omega,\theta)+
\gamma_4\,\omega^3+\CO(\omega^4), \nonumber \\
A_6(\omega,\theta)&=& A_6^B(\omega,\theta)+
\gamma_3\,\omega^3+\CO(\omega^4).
\eey

We now introduce the fixed-$t$ dispersion relations
for the non-Born parts of the structure functions $A_i^{nB}(\omega,\theta)$.
Then, with use of the dispersion relations the polarizabilities are
calculated from the imaginary parts of the non-Born terms of the
$A_i^{nB}$ and the anomalous term as follows:
\bey
\alpha+\beta
&=&{2\over \pi}\int_{\omega_{th}}^\infty{{\rm d}\omega
   \over\omega^3}\,\Im A_1^{nB}(\omega,0), \nonumber \\
\beta
&=&-{2\over \pi}\int_{\omega_{th}}^\infty{{\rm d}\omega
   \over\omega^3}\,\Im A_2^{nB}(\omega,0), \nonumber \\
\gamma_0
&=&{2\over \pi}\int_{\omega_{th}}^\infty{{\rm d}\omega
   \over\omega^4}\,\Im A_3^{nB}(\omega,0), \nonumber \\
\gamma_2
&=&{2\over \pi}\int_{\omega_{th}}^\infty{{\rm d}\omega
   \over\omega^4}\,\Im A_4^{nB}(\omega,0), \nonumber \\
\gamma_3
&=&{2\over \pi}\int_{\omega_{th}}^\infty {{\rm d}\omega
   \over\omega^4}\,\Im A_6^{nB}(\omega,0)+{\mathrm{anomalous\,\, term}},
   \nonumber \\
\gamma_4
&=&{2\over \pi}\int_{\omega_{th}}^\infty{{\rm d}\omega
   \over\omega^4}\,\Im A_5^{nB}(\omega,0)+{\mathrm{ anomalous\,\, term}}.
\label{eq:drpol}
\eey
The forward spin-polarizability $\gamma_0$ is defined by
$\gamma_0=\gamma_1-\gamma_2-2\gamma_4$.
It can be shown that the anomalous term does not contribute to
the forward spin-polarizability $\gamma_0$, and to $\gamma_2$.

The imaginary part of the scattering amplitude
is calculated from the photoabsorption amplitude of the nucleon
using the unitarity condition.
We consider the $N\pi$ and $\Delta\pi$ channels for the intermediate
states.
In a nonrelativistic approximation, the unitarity condition leads to
\ben
\Im f_{N\gamma\rightarrow N'\gamma'}
={q \over 4\pi}\sum_{B=N,\Delta}
\int{\rm d}\Omega_q f_{N'\gamma'\rightarrow B\pi}^\dagger
                    f_{N\gamma\rightarrow B\pi}, \label{eq:img}
\een
where $q$ is the pion momentum, and
the integrals are over the angle of the pion momentum.

The anomalous contribution to the polarizabilities
 is calculated in section \ref{sec:sec4}.

\section{The non-Born Amplitudes}
\label{sec:sec3}

We calculate the pion photoabsorption amplitudes in terms of the
$\gamma+N\rightarrow \pi+N$ and $\gamma+N\rightarrow \pi+\Delta$
processes in the chiral soliton model. We have shown \cite{Saito95}
that the electric and magnetic Born parts of the amplitude obtained by the
chiral soliton model satisfy the low-energy theorem
at the threshold except for the order $(m_\pi/M)^2$ term which
was recently introduced by the effect of chiral loops\cite{Kaiser92}.

\subsection{$N\pi$ channel}

At first, we evaluate the contribution of the
 $\gamma+N\rightarrow \pi+N$ channel.
The pion photoproduction amplitude is decomposed into three terms  as
\begin{equation}
f^a_{N}
=i\epsilon_{a3b}\tau^bf^{(-)}_{N}+\tau^af^{(0)}_{N}
+\delta_{a3}f^{(+)}_{N} \label{eq:isoNpi},
\een
where $\tau^a$'s are the isospin matrices, and each of the amplitudes
$f^{(\pm,0)}$ is given by the sum of the electric and magnetic parts.

The electric Born amplitude for the $\gamma+N\rightarrow \pi+N$ process
is given as
\ben
f_{N,e}^{(-)}= \left( {eG_{NN\pi}\over8\pi M} \right)
\left\{i\vect\sigma\!\cdot\!\vect\epsilon
+2{i\vect\sigma\!\cdot(\vect k-\vect q)(\vect\epsilon\!\cdot\!\vect q)
\over m_\pi^2-(k-q)^2}\right\}, \label{eq:fEN}
\een
where $k$ and $q$ are the incident photon and
the outgoing pion $4$-momenta, respectively,  $\vect\epsilon$
 the polarization vector of the incident photon,
 and $G_{NN\pi}$ the $NN\pi$ coupling constant.
In\ Fig.~\ref{fig1} we show the Born graphs of the pion photoproduction
represented in Eq.~(\ref{eq:fEN}):
1(a) and (b) are the Kroll-Ruderman  and the pion-pole terms, respectively.
We see that  $f_{N,e}^{(-)}$ is of $\CO(N_c^{1/2})$,
while  $f_{N,e}^{(+,0)}$ are of $\CO(N_c^{-1/2})$
and behave as $\CO(\omega)$.
Therefore, the latter amplitudes do not lead to finite results
without unitarization of them, and are neglected in the following.

We note that the amplitude in Eq.~(\ref{eq:fEN}) satisfies
the low-energy theorem,
and is, therefore, model-independent.
The same amplitude is known to be obtained in the leading part of the
amplitude
calculated in  HBChPT.
Furthermore, we should note that the following relativistic amplitude
\cite{Drechsel92} reduces to
Eq.~(\ref{eq:fEN}) in nonrelativistic approximation if we neglect
the $t$-dependence of $G_A(t)$:
\ben
f=\left({e G_{NN\pi} \over 8\pi M}\right)
{[\tau_a,\tau_0] \over 2} {G_A(t) \over G_A(0)}
 \bar u_{p'}\left\{\slash\epsilon+{2M \over t-\mpi^2}
 \epsilon\!\cdot\!(2q-k)\right\}\gamma_5 u_p. \label{eq:relaNpiE}
\een

The magnetic Born amplitude is given as
\begin{eqnarray}
f_{N,m}^{(-)} &=& \left( {eG_{NN\pi}\over8\pi M} \right)
\left({\mu_V\over 2M}\right)
 \left\{ -{(\vect\sigma\!\cdot\!\vect{q})
        (\vect{\sigma}\!\cdot\!\vect{s})\over\omega}
-{(\vect{\sigma}\!\cdot\!\vect{s})
  (\vect{\sigma}\!\cdot\!\vect{q})\over\omega}\right.
\nonumber \\
&&\quad{}+\left.
{1 \over 2}{\left[3\vect{s}\!\cdot\!\vect{q}
-(\vect{\sigma}\!\cdot\!{\vect{q}})
(\vect{\sigma}\!\cdot\!\vect{s})\right]
\over\omega-\Delta}
+{1 \over 2}{\left[3\vect{s}\!\cdot\!\vect{q}
-(\vect{\sigma}\!\cdot\!\vect{s})
(\vect{\sigma}\!\cdot\!\vect{q})\right]
\over\omega+\Delta}
\right\}, \nonumber \\ 
f_{N,m}^{(+)}
&=& \left( {eG_{NN\pi}\over8\pi M} \right)
\left({\mu_V\over 2M}\right)
 \left\{ -{(\vect{\sigma}\!\cdot\!\vect{q})
 (\vect{\sigma}\!\cdot\!\vect{s})\over\omega}
+{(\vect{\sigma}\!\cdot\!\vect{s})(\vect{\sigma}\!\cdot\!\vect{q})
\over\omega}\right. \nonumber \\
&&\quad{}-\left.
{\left[3\vect{s}\!\cdot\!\vect{q}-(\vect{\sigma}\!\cdot\!\vect{q})
(\vect{\sigma}\!\cdot\!\vect{s})\right]
\over\omega-\Delta}
+{\left[3\vect{s}\!\cdot\!\vect{q}-(\vect{\sigma}\!\cdot\!\vect{s})
(\vect{\sigma}\!\cdot\!\vect{q})\right]
\over\omega+\Delta}
\right\}, \label{eq:Nm}
\end{eqnarray}
where $\vect{s}=\vect{k}\times\vect{\epsilon}$,
$\Delta$ is the mass difference of the nucleon and the $\Delta(1232)$,
and $\mu_V$ is the vector part of the nucleon magnetic moment defined
by $(\mu_p-\mu_n)/2$ in units of the nuclear magneton.
Note that we introduced the nucleon- and $\Delta$-pole terms, and
 used the relation $\mu_V^{\Delta N}=-({3/\sqrt{2}})\mu_V$
in the chiral soliton model,
which is known to be correct in  large $N_c$ limit\cite{Dashen94}.
In Fig.~\ref{fig2} are shown the Born graphs in Eqs.~(\ref{eq:Nm}).
We also see that the
$f_{N,m}^{(\pm)}$ reduces to $O(N_c^{1/2})$
by the cancellation among the $N$- and $\Delta$-pole terms.
The amplitude $f_{N,m}^{(0)}$ is of $O(N_c^{-1/2})$ and
is neglected in the following.
We rewrite the $f_{N,m}^{(\pm)}$ as
\begin{equation}
f_{N,m}^{(\pm)}=\left( {eG_{NN\pi}\over8\pi M} \right)
\mu_V
\left\{ t_1^{(\pm)} P_1( \hat{\vect{q}}, \hat{\vect{s}})+
t_3^{(\pm)} P_3(\hat{\vect{q}}, \hat{\vect{s}}) \right\},
\end{equation}
where $P_1(\hat{\vect{q}},\hat{\vect{s}})
=(\vect{\sigma}\!\cdot\!\hat{\vect{q}})(\vect{\sigma}\!\cdot\!\hat{\vect{s}}
)$ and
$P_3(\hat{\vect{q}},\hat{\vect{s}})=3(\hat{\vect{q}}\!\cdot\!\hat{\vect{s}})
-
(\vect{\sigma}\!\cdot\!\hat{\vect{q}})(\vect{\sigma}\!\cdot\!\hat{\vect{s}})
$
are the $P$-wave projection operators for the $J=1/2$ and $J=3/2$ states,
respectively, and $\hat{\vect{q}}=\vect{q}/q$ and
$\hat{\vect{s}}=\vect{s}/k$.
We obtain
\begin{eqnarray}
t_1^{(+)} &=& 2t_1^{(-)} =
-{2q\over3M}{\Delta \over \omega+\Delta}, \nonumber\\
t_3^{(+)} &=& {q\omega\over2M}
\left[-{2\Delta \over \omega^2-\Delta^2+i\Delta \Gamma_\Delta}
+{2\over3}{\Delta \over \omega(\omega+\Delta)}\right], \nonumber\\
t_3^{(-)} &=& {q\omega\over2M}
\left[{\Delta \over \omega^2-\Delta^2+i\Delta \Gamma_\Delta}
-{2\over3}{\Delta \over \omega(\omega+\Delta)}\right].
\label{eq:deltapole}
\end{eqnarray}
Here, in order to avoid divergent dispersion integral due to the pole
at $\omega=\Delta$,
we have introduced the finite width of the $\Delta$ state given by
\begin{equation}
\Gamma_\Delta={1\over6\pi}\left({G_{\Delta N\pi}\over 2M}\right)^2 q^3
\label{eq:width}
\end{equation}
with $G_{\Delta N\pi}=-(3/\sqrt{2})G_{NN\pi}$, whose relation
is also correct in
large $N_c$ limit. This is the expression
given by Kokkedee without  relativistic correction\cite{Adkins}, and
yields $145{\rm MeV}$ with the experimental value of $G_{NN\pi}$
at $q=227$MeV.

The amplitudes $f_{N,e}^{(-)}$ and $f_{N,m}^{(\pm)}$ survive
in the leading order of $1/N_c$ expansion,
so that we find from Eq.~(\ref{eq:isoNpi})
\ben
f_N^{\dagger} f_N
=2f_{N,e}^{(-)\dagger}f_{N,e}^{(-)}
+\(2f_{N,m}^{(-)\dagger}f_{N,m}^{(-)}+f_{N,m}^{(+)\dagger}f_{N,m}^{(+)}\)
+2\(f_{N,e}^{(-)\dagger}f_{N,m}^{(-)}+f_{N,m}^{(-)\dagger}f_{N,e}^{(-)}\).
\een
We call the first,  the second and the third terms as the electric,
magnetic and interference parts, respectively.

We now begin with the contribution of the electric part for calculating
the imaginary part of the Compton scattering amplitude in use of
the unitarity condition in Eq.~(\ref{eq:img}).
Using the integral formulas in Appendix \ref{sec:integral}, we obtain
\bey
\int{\rm d}\Omega_q\, 2f_{N,e}^{(-)\dagger} f_{N,e}^{(-)}
&=&8\pi \({eG_{NN\pi} \over 8\pi M}\)^2 \nonumber \\
&&\quad\times \left\{\[1-2I_{2}(v)+(\hat{\vect k}'\!\cdot\!\hat{\vect
k}+v^2)J_{1}(v,\theta)
    -2(1+\hat{\vect k}'\!\cdot\!\hat{\vect k})J_{6}(v,\theta)\]
    \vect\epsilon^{\prime *}\!\cdot\!\vect\epsilon \right. \nonumber \\
&&\qquad+\[(\hat{\vect k}'\!\cdot\!\hat{\vect k}+v^2)J_{3}(v,\theta)
    -2(1+\hat{\vect k}'\!\cdot\!\hat{\vect
k})J_{5}(v,\theta)-2J_{6}(v,\theta)\]
   (\vect\epsilon^{\prime *}\!\cdot\!\hat{\vect k})
   (\vect\epsilon\!\cdot\!\hat{\vect k}') \nonumber \\
&&\qquad+(1-2I_{2}(v))i\vect\sigma\!\cdot\!
     \vect\epsilon^{\prime *}\!\times\!\vect\epsilon \nonumber \\
&&\qquad+(J_{1}(v,\theta)-2J_{6}(v,\theta))i\vect\sigma\!\cdot\!
     (\hat{\vect k}'\!\times\hat{\vect k})
     (\vect\epsilon^{\prime *}\!\cdot\!\vect\epsilon) \nonumber \\
&&\qquad-J_{6}(v,\theta)\,i\vect\sigma\!\cdot\!
   \[(\vect\epsilon^{\prime *}\!\times\hat{\vect k})
     (\vect\epsilon\!\cdot\!\hat{\vect k}')
    -(\vect\epsilon\!\times\hat{\vect k}')
     (\vect\epsilon^{\prime *}\!\cdot\!\hat{\vect k})\] \nonumber \\
&&\qquad+J_{6}(v,\theta)\,i\vect\sigma\!\cdot\!
   \[(\vect\epsilon^{\prime *}\!\times\hat{\vect k}')
     (\vect\epsilon\!\cdot\!\hat{\vect k}')
    -(\vect\epsilon\!\times\hat{\vect k})
     (\vect\epsilon^{\prime *}\!\cdot\!\hat{\vect k})\] \nonumber \\
&&\qquad\left.+(J_{3}(v,\theta)-2J_{5}(v,\theta))
    i\vect\sigma\!\cdot\!(\hat{\vect k}'\!\times\!\hat{\vect k})
   (\vect\epsilon^{\prime *}\!\cdot\!\hat{\vect k})
   (\vect\epsilon\!\cdot\!\hat{\vect k}')\right\}, \label{eq:absfEN}
\eey
where $\theta$ is the scattering angle in the center-of-mass system,
and $I_i(v)$ for $i=1,\ldots,5$ and $J_j(v,\theta)$ for $j=1,\ldots,6$
with $v$ the pion velocity are defined in Appendix A.
It is known that
the spin factor of the last term in Eq.~(\ref{eq:absfEN}),
$i\vect\sigma\!\cdot\!(\hat{\vect k}'\!\times\!\hat{\vect k})
(\vect\epsilon^{\prime *}\!\cdot\!\hat{\vect k})
(\vect\epsilon\!\cdot\!\hat{\vect k}')$,
is not independent of  the other spin factors; actually, we
have
\bey
i\vect\sigma\!\cdot\!(\hat{\vect k}'\!\times\!\hat{\vect k})
(\vect\epsilon^{\prime *}\!\cdot\!\hat{\vect k})
(\vect\epsilon\!\cdot\!\hat{\vect k}')
&=&\( 1-( \hat{\vect k}' \cdot \hat{\vect k} )^2 \)
i\vect\sigma\!\cdot\!\vect\epsilon^{\prime *}
\!\times\! \vect\epsilon
+ (\hat{\vect k}'\cdot\hat{\vect k})\,i\vect\sigma\!\cdot\!
   \[(\vect\epsilon^{\prime *}\!\times\hat{\vect k})
     (\vect\epsilon\!\cdot\!\hat{\vect k}')
    -(\vect\epsilon\!\times\hat{\vect k}')
     (\vect\epsilon^{\prime *}\!\cdot\!\hat{\vect k})\] \nonumber \\
&&\qquad - i\vect\sigma\!\cdot\!
   \[(\vect\epsilon^{\prime *}\!\times\hat{\vect k}')
     (\vect\epsilon\!\cdot\!\hat{\vect k}')
    -(\vect\epsilon\!\times\hat{\vect k})
     (\vect\epsilon^{\prime *}\!\cdot\!\hat{\vect k})\] . \label{eq:gamma7}
\eey
The last term is, therefore, redistributed into the
third, the fifth and the sixth terms in Eq.~(\ref{eq:absfEN}),
and it turns out that
$\Im A_5(\omega,0)=\Im A_6(\omega,0)=0$ due to the cancellations;
namely, we can read off from Eqs.~(\ref{eq:absfEN}) and (\ref{eq:gamma7})
that $\Im A_5$ and $\Im A_6$ at $\theta=0$ are proportional to
$J_3(v,0)-2 J_5(v,0)-J_6(v,0)$, but this is identically zero,
as seen from the definitions in Appendix A.
Therefore, the dispersion integrals in Eqs.~(\ref{eq:drpol})
yield $\gamma_3^{N,e}=\gamma_4^{N,e}=0$, except for the anomalous term.
This is, however, not correct:
The spin-factor of the last term  in Eq.~(\ref{eq:absfEN})
originally consists of a fourth-order product of the photon momenta
as $i\vect\sigma\!\cdot\!({\vect k}'\!\times\!{\vect k})
(\vect\epsilon^{\prime *}\!\cdot\!{\vect k})
(\vect\epsilon\!\cdot\!{\vect k}')$,
because the real part of the amplitude must be an analytic function
around at $\omega=0$ but the unit vectors
$\hat{\vect k}$ and $\hat{\vect k}'$ are not analytic.
Consequently, the real part of the amplitude as a coefficient of
the last spin-factor should be of $\CO(\omega^4)$ or higher.
Note that there is no possibility
for the real part of the amplitude
to have an extra factor $1/\omega^2$ to reduce the power
down to $\CO(\omega^3)$ at low energies.
We shall explicitly show that this argument is correct, by comparing it
with  Compton scattering amplitude
 in HBChPT in Appendix \ref{sec:NE}.
Therefore, we do not include the contributions from the last term
of Eq.~(\ref{eq:absfEN}) into the dispersion integrals.\footnote
{L'vov already argued about this problem
and states that the dispersion relation does not work in this case
from the high-energy behavior of the relativistic invariant amplitudes
with Regge-pole assumption\cite{Lvov98}.}

We thus obtain the imaginary parts of the structure functions,
 $\Im A_i^{N,e}(\omega,\theta)$, for the electric part in the $N\pi$ channel
as
\bey
\Im A_1^{N,e}(\omega,\theta)
&=&{e^2\over 4\pi}{G_{NN\pi}^2\over 8\pi M^2}
  \,q\[1-2I_{2}(v)+(\cos\theta+v^2)J_{1}(v,\theta)-2(1+\cos\theta)
  J_{6}(v,\theta)\], \nonumber \\
\Im A_2^{N,e}(\omega,\theta)
&=&{e^2\over 4\pi}{G_{NN\pi}^2\over 8\pi M^2}
  \,q\[(\cos\theta+v^2)J_{3}(v,\theta)-2(1+\cos\theta)J_{5}(v,\theta)
  -2J_{6}(v,\theta)\], \nonumber \\
\Im A_3^{N,e}(\omega,\theta)
&=&{e^2\over 4\pi}{G_{NN\pi}^2\over 8\pi M^2}
  \,q\(1-2I_{2}(v)\), \nonumber \\
\Im A_4^{N,e}(\omega,\theta)
&=&{e^2\over 4\pi}{G_{NN\pi}^2\over 8\pi M^2}
  \,q\(J_{1}(v,\theta)-2J_{6}(v,\theta)\), \nonumber \\
\Im A_5^{N,e}(\omega,\theta)
&=&-A_6^{N,e}(\omega,\theta)
  =-{e^2\over 4\pi}{G_{NN\pi}^2\over 8\pi M^2}
  \,qJ_{6}(v,\theta). \label{eq:A5EN}
\eey
Using the imaginary parts we integrate the dispersion integrals
in Eqs.~(\ref{eq:drpol}),
and obtain the contribution of the electric part to
the polarizabilities:
\bey
\alpha^{N,e}  &=&{e^2\over 4\pi}{10 G_{NN\pi}^2\over 192\pi M^2\mpi},\quad
\beta^{N,e}={1\over10}\, \alpha^{N,e}, \nonumber \\
\gamma^{N,e}_0&=&{e^2\over 4\pi}{G_{NN\pi}^2\over 24\pi^2 M^2\mpi^2},
\nonumber \\
\gamma_1^{N,e}&=&\gamma^{N,e}_0,\quad
\gamma_2^{N,e}={1\over 2}\gamma^{N,e}_0,\quad
\gamma_3^{N,e}={1\over 4}\gamma^{N,e}_0,\quad
\gamma_4^{N,e}=-{1\over 4}\gamma^{N,e}_0 .
\eey
In terms of the Goldberger-Treiman relation,
the results are shown to be the same as those of the $N\pi$-loops
 in HBChPT\cite{Hemmert98}.
The $1/m_\pi$ and $1/m_\pi^2$ dependences mean that these are
the contributions from the pion cloud.
Because the proton-neutron difference
depends on the amplitude $f_{N,e}^{(0)}$,
we predict only the average between them.
It is also known that the prediction of  HBChPT up to chiral order
$\epsilon^3$ yields no isospin dependence.

For the the magnetic part we obtain the imaginary parts,
$\Im A_i^{N,m}(\omega,\theta)$, in the same way as those in the electric
part
\bey
\Im A_2^{N,m}(\omega,\theta)
&=&-{e^2\over 4\pi}{G_{NN\pi}^2\over 16\pi M^2}\,\mu_V^2\,  q
  \[2\(\abs{t_1^{(-)}}^2+2\abs{t_3^{(-)}}^2\)
    +\(\abs{t_1^{(+)}}^2+2\abs{t_3^{(+)}}^2\)\], \nonumber \\
\Im A_1^{N,m}(\omega,\theta)
&=&-\Im A_2^{N,m}(\omega,\theta)\cos\theta, \nonumber \\
\Im A_4^{N,m}(\omega,\theta)
&=&{e^2\over 4\pi}{G_{NN\pi}^2\over 16\pi M^2}\,\mu_V^2\,  q
  \[2\(\abs{t_1^{(-)}}^2-\abs{t_3^{(-)}}^2\)
    +\(\abs{t_1^{(+)}}^2-\abs{t_3^{(+)}}^2\)\], \nonumber \\
\Im A_3^{N,m}(\omega,\theta)
&=& \Im A_4^{N,m}(\omega,\theta)\cos\theta, \nonumber \\
\Im A_5^{N,m}(\omega,\theta) &=& -\Im A_4^{N,m}(\omega,\theta), \nonumber \\
\Im A_6^{N,m}(\omega,\theta) &=&0.
\eey
Substituting these terms into the dispersion integrals we find
\ben
\gamma_2^{N,m}=-\gamma_4^{N,m}=\gamma^{N,m}_0, \quad
\alpha^{N,m}=\gamma_1^{N,m}=\gamma_3^{N,m}=0, \label{eq:polM}
\een
which are equal to the results of the $\Delta$-pole contribution
in HBChPT\cite{Hemmert98}.
At the limit of $\Gamma_\Delta\rightarrow 0$
we obtain
\ben
\beta^{N,m}={e^2\over 4\pi}{\mu_V^2\over M^2}{1\over\Delta}, \quad
\gamma^{N,m}_0=-{e^2\over 4\pi}{\mu_V^2\over 2M^2}{1\over\Delta^2}.
\een
Identifying the coefficient of a counter term in the HBChPT,
 $b_1$, as $\mu_V^{\Delta N}/2$
we see that this is just the
$\Delta$-pole contribution in  HBChPT\cite{Hemmert97}.
$b_1$ is numerically about $-2.5\pm0.35$
from a tree level relativistic analysis
\footnote{Recently, Hemmert \etal\  used the value $b_1^2=3.85\pm 0.15$
obtained within the ``small scale expansion''\cite{Hemmert98}.
The numerical results using this value are
similar to our results using the finite $\Delta$ width.
},
while $\mu_V^{\Delta N}/2$ is
$-2.5$ with use of the experimental value of $\mu_V$.

The interference part of the electric and magnetic terms is
given as follows, by noting that $t_3^{(-)}$ are complex functions due to
the $\Delta$ width:
\bey
\lefteqn{\int{\rm d}\Omega_q\,
 2\(f_{N,e}^{(-)\dagger} f_{N,m}^{(-)}
+f_{N,m}^{(-)\dagger} f_{N,e}^{(-)}\)} \nonumber \\
&=& 8\pi \({eG_{NN\pi} \over 8\pi M}\)^2\mu_V{1\over v} \nonumber \\
&&\quad\times\left\{
   2I_{2}(v)\(t_1^{(-)}-\Re t_3^{(-)}\)
    \[(\vect\epsilon^{\prime *}\!\cdot\!\vect\epsilon)
      (\hat{\vect k'}\!\cdot\!\hat{\vect k})
     -(\vect\epsilon^{\prime *}\!\cdot\!\hat{\vect k})
      (\vect\epsilon\!\cdot\!\hat{\vect k'})\] \right. \nonumber \\
&&\qquad+2\[I_{2}(v)t_1^{(-)}+(2I_{2}(v)-3I_{4}(v))\Re t_3^{-}\]
   i\vect\sigma\!\cdot\!
   (\vect\epsilon^{\prime *}\!\times\!\vect\epsilon)
   (\hat{\vect k'}\!\cdot\!\hat{\vect k}) \nonumber \\
&&\qquad+2\[I_{2}(v)t_1^{(-)}-(I_{2}(v)-3I_{4}(v))\Re t_3^{(-)}\]
   i\vect\sigma\!\cdot\!
   (\hat{\vect k}'\!\times\!\hat{\vect k})
   (\vect\epsilon^{\prime *}\!\cdot\!\vect\epsilon) \nonumber \\
&&\qquad-I_{2}(v)\(2t_1^{(-)}+\Re t_3^{(-)}\)
   i\vect\sigma\!\cdot\!
   \[(\vect\epsilon^{\prime *}\!\times\!\hat{\vect k})
     (\vect\epsilon\!\cdot\!\hat{\vect k}')
    -(\vect\epsilon\!\times\!\hat{\vect k}')
     (\vect\epsilon^{\prime *}\!\cdot\!\hat{\vect k})\] \nonumber \\
&&\qquad\left. +(3I_{2}(v)+6I_{4}(v))\Im t_3^{(-)}
   i\vect\sigma\!\cdot\!
   \[(\vect\epsilon^{\prime *}\!\times\!\hat{\vect k})
     (\vect\epsilon\!\cdot\!\hat{\vect k}')
    +(\vect\epsilon\!\times\!\hat{\vect k}')
     (\vect\epsilon^{\prime *}\!\cdot\!\hat{\vect k})\] \right\}.
\eey
The last term, which is anti-hermite, disappears at the narrow width
limit of the $\Delta$ state and is neglected in the following.
We then get
\bey
\Im A_2^{N,i}(\omega,\theta)
&=&-{e^2\over 4\pi}{G_{NN\pi}^2\over 8\pi M^2}\,\mu_V{2q\over v}
   I_{2}(v)\(t_1^{(-)}-\Re t_3^{(-)}\), \nonumber \\
\Im A_1^{N,i}(\omega,\theta)
&=&-\Im A_2^{N,i}(\omega,\theta)\cos\theta, \nonumber \\
\Im A_3^{N,i}(\omega,\theta)
&=&{e^2\over 4\pi}{G_{NN\pi}^2\over 8\pi M^2}\,\mu_V{2q\over v}
   \[I_{2}(v)t_1^{(-)}+(2I_{2}(v)-3I_{4}(v))\Re t_3^{(-)}\] \cos\theta, 
   \nonumber \\
\Im A_4^{N,i}(\omega,\theta)
&=&{e^2\over 4\pi}{G_{NN\pi}^2\over 8\pi M^2}\,\mu_V{2q\over v}
   \[I_{2}(v)t_1^{(-)}-(I_{2}(v)-3I_{4}(v))\Re t_3^{(-)}\], \nonumber \\
\Im A_5^{N,i}(\omega,\theta)
&=&-{e^2\over 4\pi}{G_{NN\pi}^2\over 8\pi M^2}\,\mu_V{q\over v}
   I_{2}(v)\(2t_1^{(-)}+\Re t_3^{(-)}\), \nonumber \\
\Im A_6^{N,i}(\omega,\theta)
&=&0.
\eey
From these equations, we find
\ben
\alpha^{N,i}=\gamma_1^{N,i}=\gamma_3^{N,i}=0.
\label{eq:polEM}
\een
As an example we show a contribution to  Compton scattering  from the
interference part in Fig.~\ref{fig3} diagrammatically.
This kind of diagrams is not taken into account in HBChPT.

\subsection{$\Delta\pi$ channel}

Next we examine the $\gamma+N\rightarrow \pi+\Delta$ contribution.
The amplitude is also decomposed into three terms as
\begin{equation}
f^a_{\Delta}=
i\epsilon_{a3b}\CT^b f^{(-)}_{\Delta}+\CT^a f^{(0)}_{\Delta}
+\CT_{a3}^+f^{(+)}_{\Delta}, \label{eq:isoDpi}
\end{equation}
where $\CT^a$ is the transition isospin matrix from $N$ to
$\Delta$, and $\CT_{a3}^+=\CT^a\half
\tau^3+\half\CT^3_{\Delta\Delta}\CT^a$.
The electric part is obtained by replacing
$\vect{\sigma}$ and $G_{NN\pi}$ in Eq.~(\ref{eq:fEN})
by the transition spin operator $\vect{S}_{\Delta N}$ and
$G_{\Delta N\pi}$, respectively.
The magnetic part is given by
\begin{eqnarray}
f_{\Delta,m}^{(-)} &=&
\left( {eG_{\Delta N\pi}\over8\pi M} \right)
\left({\mu_V\over2M}\right)
 \left\{ -{(\vect{ S}_{\Delta N}\!\cdot\!\vect{ q})
        (\vect{\sigma}\!\cdot\!{\vect{s}})\over\omega}
-{4\over5}{(\vect{ S}_{\Delta\Delta}\!\cdot\!\vect{ q})
(\vect{ S}_{\Delta N}\!\cdot\!\vect{ s})\over\omega_q}\right. \nonumber\\
&&\quad{}+\left.
2{(\vect{ S}_{\Delta N}\!\cdot\!\vect{s})(\vect{\sigma}\!\cdot\!\vect{q})
  \over\omega_q}
-{1\over5}{(\vect{S}_{\Delta\Delta}\!\cdot\!\vect{s})
           (\vect{S}_{\Delta N}\!\cdot\!\vect{q})\over\omega}
\right\}, \nonumber \\
f_{\Delta,m}^{(+)} &=&
\left( {eG_{\Delta N\pi}\over8\pi M} \right)
\left({\mu_V\over2M}\right)
 \left\{ -{(\vect{S}_{\Delta N}\!\cdot\!\vect{q})
           (\vect\sigma\!\cdot\!\vect{s})\over\omega}
-{1\over5}{(\vect{S}_{\Delta\Delta}\!\cdot\!\vect{q})
(\vect{S}_{\Delta N}\!\cdot\!\vect{s})\over\omega_q}\right. \nonumber\\
&&\quad{}+\left.
{(\vect{S}_{\Delta N}\!\cdot\!\vect{s})
 (\vect{\sigma}\!\cdot\!\vect{q})\over\omega_q}
+{1\over5}{(\vect{S}_{\Delta\Delta}\!\cdot\!\vect{s})
           (\vect{S}_{\Delta N}\!\cdot\!\vect{q})\over\omega}
\right\},
\end{eqnarray}
where $\vect{S}_{\Delta\Delta}$ is the spin matrix for the $\Delta$ state,
and $\omega_q$ is the energy of pion.
From Eq.~(\ref{eq:isoDpi}) we get in the leading order of
$1/N_c$ expansion
\bey
f_\Delta^{\dagger} f_\Delta
&=&{4\over 3}f_{\Delta,e}^{(-)\dagger}f_{\Delta,e}^{(-)}
+\[{4\over 3}f_{\Delta,m}^{(-)\dagger}f_{\Delta,m}^{(-)}
             -2\(f_{\Delta,m}^{(-)\dagger}f_{\Delta,m}^{(+)}
+f_{\Delta,m}^{(+)\dagger}f_{\Delta,m}^{(-)}\)
     +{14\over 3}f_{\Delta,m}^{(+)\dagger}f_{\Delta,m}^{(+)}\] \nonumber \\
&&+\[{4\over 3}\(f_{\Delta,e}^{(-)\dagger}f_{\Delta,m}^{(-)}
+f{\Delta,m}^{(-)\dagger}f_{\Delta,e}^{(-)}\)
             -2\(f_{\Delta,e}^{(-)\dagger}f_{\Delta,m}^{(+)}
+f_{\Delta,m}^{(+)\dagger}f_{\Delta,e}^{(-)}\)\].
\eey

The results are as follows:
The electric contributions are given by
\bey
\Im A_1^{\Delta,e}(\omega,\theta)
&=&{e^2 \over 4\pi}{G_{\Delta N\pi}^2\over 18\pi M^2}\,
   {q\over b}\[b-2I_{2}(v)
+b \(\cos\theta+{v^2\over b^2}\)J_{1}(v,\theta)
       -2(1+\cos\theta)J_{6}(v,\theta)\], \nonumber \\
\Im A_2^{\Delta,e}(\omega,\theta)
&=&{e^2 \over 4\pi}{G_{\Delta N\pi}^2\over 18\pi M^2}\,
   {q\over b}\[b \(\cos\theta+{v^2\over b^2}\)J_{3}(v,\theta)
       -2\(1+\cos\theta\)J_{5}(v,\theta)
-2J_{6}(v,\theta) \], \nonumber \\
\Im A_3^{\Delta,e}(\omega,\theta)
&=&-{e^2 \over 4\pi}{G_{\Delta N\pi}^2\over 36\pi M^2}\,
   {q\over b}\[b-2I_{2}(v)\], \nonumber \\
\Im A_4^{\Delta,e}(\omega,\theta)
&=&-{e^2 \over 4\pi}{G_{\Delta N\pi}^2\over 36\pi M^2}\,
   {q\over b}\[b J_{1}(v,\theta)-2J_{6}(v,\theta)\], \nonumber \\
\Im A_5^{\Delta,e}(\omega,\theta)
&=& -\Im A_6^{\Delta,e}(\omega,\theta)=
{e^2 \over 4\pi}{G_{\Delta N\pi}^2\over 36\pi M^2}\,
   {q\over b}J_{6}(v,\theta),
\eey
where $b=\omega/\omega_q$.\footnote{In a previous paper\cite{Tanushi},
we have not taken into account the energy transfer in the pion propagator
in Eq.~(\ref{eq:fEN}), because the energy transfer is of higher order in
the $1/N_c$ expansion. This leads to  different expressions to the
amplitudes depending on the parameter $a=(1+b^2)/2$ as shown in the paper.
However, we note that $a=b+\CO(1/N_c^2)$.
Then, if we put $a=b$ by neglecting
the $1/N_c^2$ term, we obtain the same results as those by
the pion propagator with the energy transfer.}
The dispersion integrals with the above amplitudes  yield
\bey
\alpha^{\Delta,e}
&=& {e^2\over4\pi} {G^2_{\Delta N\pi}\over 216\pi^2M^2}
\[
 {9\Delta\over \Delta^2-m_\pi^2}+{\Delta\over m_\pi^2}
+{\Delta^2-10m_\pi^2\over(\Delta^2-m_\pi^2)^{3/2}}\ln R
\],   \nonumber \\
\beta^{\Delta,e}
&=& {e^2\over4\pi} {G^2_{\Delta N\pi}\over 216\pi^2M^2}
\[
 -{\Delta\over m_\pi^2}
+{\ln R\over(\Delta^2-m_\pi^2)^{1/2}}
\],   \nonumber \\
\gamma_1^{\Delta,e}
&=& -{e^2\over4\pi} {G^2_{\Delta N\pi}\over 216\pi^2M^2}
\[ {\Delta^2+2m_\pi^2\over (\Delta^2-m_\pi^2)^2}
   -{3m_\pi^2\Delta \ln R\over(\Delta^2-m_\pi^2)^{5/2}}\],  \nonumber \\
\gamma_2^{\Delta,e}
&=& -{e^2\over4\pi} {G^2_{\Delta N\pi}\over 216\pi^2M^2}
\[ {1\over \Delta^2-m_\pi^2}
   -{\Delta \ln R\over(\Delta^2-m_\pi^2)^{3/2}}\], \nonumber \\
\gamma_3^{\Delta,e} &=&-\gamma_4^{\Delta,e}={1\over2} \gamma_2^{\Delta,e}
\eey
with
\ben
R={\Delta\over m_\pi}+\sqrt{{\Delta^2\over m_\pi^2}-1}.
\een
The results for the spin-polarizabilities are the same as the results of
the $\Delta\pi$-loops in HBChPT, but there are a little difference for
the spin-independent polarizabilities: we see that the sum
$\alpha+\beta$ is the same as that in \cite{Hemmert97}, but
there is no $\Delta/m_\pi^2$ term in \cite{Hemmert97}.
The magnetic terms are
\bey
\Im A_2^{\Delta,m}(\omega,\theta)
&=&-{e^2 \over 4\pi}{G_{\Delta N\pi}^2\Delta^2\over 54\pi M^4}\,\mu_V^2
   \, q v^2, \nonumber \\
\Im A_1^{\Delta,m}(\omega,\theta)
&=& -\Im A_2^{\Delta,m}(\omega,\theta)\cos\theta, \nonumber \\
\Im A_3^{\Delta,m}(\omega,\theta)
&=& {1\over4}\, \Im A_1^{\Delta,m}(\omega,\theta), \nonumber \\
\Im A_4^{\Delta,m}(\omega,\theta)
&=& -\Im A_5^{\Delta,m}(\omega,\theta)=-{1\over4}\,
\Im A_2^{\Delta,m}(\omega,\theta), \nonumber \\
\Im A_6^{\Delta,m}(\omega,\theta)
&=&0.
\eey
We then obtain
\bey
\alpha^{\Delta,m} &=& 0, \nonumber \\
\beta^{\Delta,m} &=& {e^2\over4\pi}{G_{\Delta N\pi}^2 \mu_V^2
\over 54\pi^2 M^4} \left[{\Delta^2+6m_\pi^2\over \Delta}
-{3\pi m_\pi^3\over\Delta^2}
- {3m_\pi^2(\Delta^2-2m_\pi^2)\over\Delta^2\sqrt{\Delta^2-m_\pi^2}
} \ln R \right], \nonumber \\
\gamma_1^{\Delta,m}&=& \gamma_3^{\Delta,m}=0,  \nonumber \\
\gamma_2^{\Delta,m}&=& -\gamma_4^{\Delta,m}=
-{e^2\over4\pi} {G^2_{\Delta N\pi} \mu_V^2 \over 108\pi^2 M^4}
{m_\pi^3\over \Delta^3}
\left[{\Delta(24m_\pi^4-20m_\pi^2\Delta^2-\Delta^4)\over
6m_\pi^3(\Delta^2-m_\pi^2)}\right. \nonumber \\
&&\qquad\left. +2\pi
    +{(8m_\pi^4-12m_\pi^2\Delta^2+3\Delta^4) \ln R\over
2m_\pi(\Delta^2-m_\pi^2)^{3/2}}\right].
\eey
The interference part is calculated to be
\bey
\Im A_2^{\Delta,i}(\omega,\theta)
&=&{e^2 \over 4\pi}{G_{\Delta N\pi}^2\Delta\over 18\pi M^3}\,\mu_V
   q I_{2}(v), \nonumber \\
\Im A_1^{\Delta,i}(\omega,\theta)
&=&\Im A_2^{\Delta,i}(\omega,\theta)\cos\theta, \nonumber \\
\Im A_3^{\Delta,i}(\omega,\theta)
&=&{e^2 \over 4\pi}{G_{\Delta N\pi}^2\Delta\over 36\pi M^3}\,\mu_V
   {q\over b}I_{4}(v)\cos\theta, \nonumber \\
\Im A_4^{\Delta,i}(\omega,\theta)
&=&{e^2 \over 4\pi}{G_{\Delta N\pi}^2\Delta\over 36\pi M^3}\,\mu_V
   q \(I_{2}(v)-{1\over b}I_{4}(v)\), \nonumber \\
\Im A_5^{\Delta,i}(\omega,\theta)
&=& -{1\over4}\,\Im A_2^{\Delta,i}(\omega,\theta), \nonumber \\
\Im A_6^{\Delta,i}(\omega,\theta)
&=&0.
\eey
We note that the relations of the polarizabilities
(\ref{eq:polM})  are also realized
in the $\Delta\pi$ channel.

\section{Contribution of the anomalous part}
\label{sec:sec4}

The anomalous part is described as the contribution
from the Wess-Zumino-Witten term\cite{Wess}, whose Lagrangian is given by
\ben
L_{WZW}=e^2\int{\rm d}^3 x\,\epsilon^{\mu\nu\rho\sigma}
        \del_\mu A_\nu(x) A_\rho(x) W_\sigma(x)
\een
with
\ben
W_{\sigma}(x)=-{1\over 8\pi^2\fpi^2}
  \[\Phi_0(x) \del_\sigma \Phi_3(x) - \Phi_3(x) \del_\sigma \Phi_0(x)\],
\een
where $\Phi_a(x)$ and $A_\mu(x)$ are the pion and photon fields,
respectively.

Following the method in Ref.\cite{Saito95},
we obtain the anomalous part as the following seagull term:
\ben
f_{\mathrm{seagull}}
=\vect\epsilon_i'\vect\epsilon_j{i\over 4\pi}
\int {\rm d}^3 y\, e^{-i\vect{k'}\!\cdot\!\vect{y}}
\bra{N(\vect p')}\left\{[A_i(\vect{y}),\CJ_j(0)]
-i\omega'[A_i(\vect{y}),\CJ_j(0)]\right\}\ket{N(\vect{p})}.
\een
Here, the interaction current $\CJ_i$ is calculated
from the Wess-Zumino-Witten Lagrangian as
\ben
\CJ_i=2e^2\epsilon_{ijk}\pi^A_j W_k,
\een
where $\pi^A_j$ denotes the momentum field of photon.
We then obtain
\ben
f_{\mathrm{seagull}}=
2i\omega\epsilon_{ijk}\vect\epsilon_i'\vect\epsilon_j{e^2\over 4\pi}
\bra{N(\vect p')}W_k\ket{N(\vect{p})}.
\een

The pion fields between baryon states are reduced to
the classical soliton fields as follows:
$\Phi_0(x)=\hat\phi_0(\vect{x}-\vect{X}(t))$,
$\Phi_a(x)=R_{ai}\hat\phi_i(\vect{x}-\vect{X}(t))$ with
$\hat\phi_0(\vect{r})=\fpi\cos F(r)$ and
$\hat\phi_i(\vect{r})=\fpi\hat r_i\sin F(r)$,
where $F(r)$ is the profile function,
$R_{ai}$ the orthogonal rotation matrix,
and $\vect{X}(t)$  the center of the soliton.
The matrix element is then given by
\ben
\bra{N(\vect p')}W_k\ket{N(\vect{p})}
=-{1\over 8\pi^2}\Lambda_{NN}\tau_3\sigma_l{1\over \omega_q^2}
F_{lk}(\vect{q}^2),
\een
where $\Lambda_{NN}$ is defined to be $-1/3$,
$\vect{q}=\vect{k'}-\vect{k}=\vect{p}-\vect{p'}$ is the pion momentum,
and
\bey
F_{lk}(\vect{q}^2)
&=& {\omega_q^2\over f_\pi^2} \int{\rm d}^3 r\,
e^{i\vect{q}\!\cdot\!\vect{r}}
   \left[\hat\phi_0(\vect{r})\del_k\hat\phi_l(\vect{r})
    -\hat\phi_l(\vect{r})\del_k\hat\phi_0(\vect{r})\right].
\eey
$F_{lk}(\vect{q}^2)$ is calculated to be
\bey
F_{lk}(\vect{q}^2)
&=& -iq_k q_l \omega_q^2\int{\rm d}^3 r {j_1(qr) \over q}\cos F(r) \sin F(r)
+\CO(\omega_q^2) \nonumber \\
&&\longrightarrow
{q_k q_l\over\Lambda_{NN}}{G_{NN\pi}\over 2M\fpi},
\eey
where we have taken the limit of $\omega_q^2\rightarrow0$\cite{Saito93}.
Neglecting the $\vect{q}^2$ dependence of $F_{lk}(\vect{q}^2)$ we
finally obtain
\ben
f_{\mathrm{seagull}}=-{e^2\over 4\pi}{\omega \over 4\pi^2\mpi^2}
{G_{NN\pi}\over 2M\fpi}
i\vect{q}\!\cdot\!\vect\epsilon'\!\times\!\vect\epsilon
(\vect{\sigma}\!\cdot\!\vect{q})
\tau_3.
\een
In Fig.~\ref{fig4} we show the diagram of the anomalous part of
Compton scattering.

The spin-polarizabilities from the anomalous part can be read off
as follows:
\bey
\gamma_1^{\mathrm{anom}}
&=&-{e^2\over 4\pi}{G_{NN\pi} \over 4\pi^2 M\fpi\mpi^2} \tau_3, \nonumber \\
\gamma_2^{\mathrm{anom}}
&=& 0, \quad
\gamma_3^{\mathrm{anom}} = -\gamma_4^{\mathrm{anom}}
 = -{1\over2}\gamma_1^{\mathrm{anom}}.
\eey
These results are also the same as those in HBChPT.
We see that there is no contribution to the forward spin-polarizability
$\gamma_0$.

\section{Results and Discussion}
\label{sec:sec5}

Numerical results of the polarizabilities for the non-Born part
are given in Table \ref{tb:table1},
where  empirical values of  constants in the  formulas are used;
namely, $f_\pi=93$MeV, $M=939$MeV, $\Delta=293$MeV,
$G_{\pi NN}=13.5$ and $m_\pi=138$MeV.
The results of HBChPT\cite{Hemmert98} are also given in the table.
In the results of  HBChPT we give two cases for the $\Delta$-pole terms:
the upper ones are obtained using the $\pi N\Delta$ and $\gamma N \Delta$
coupling constants determined by the ``small scale expansion'' itself,
while the lower in the parentheses are
by a tree-level relativistic analysis.

We note that, although the electric part of the polarizabilities
is the same as the $N\pi$ and $\Delta\pi$ loops in  HBChPT,
the numerical values are slightly different, because
we did not used the Goldberger-Treiman relation. For the $\Delta\pi$ loops
the values in the parentheses are corresponding to the electric part
for the $\Delta\pi$ channel.

The magnetic part of the $N\pi$ channel with the narrow-width limit
in the chiral soliton model, which is shown in the parentheses,
is the same as those of the $\Delta$-poles in HBChPT.
In the chiral soliton model the finite-width effect of
the $\Delta$ particle reduces
the magnetic contribution in the same as for the
magnetic polarizability $\beta$\cite{Tanushi}.
In HBChPT the numerical results with the parameters
which are determined by the ``small scale expansion'' are similar to
the ones with the finite width in the chiral soliton model.

As shown in section \ref{sec:sec3},
no interference part of the electric and magnetic amplitudes is calculated
in HBChPT, because these terms are of higher orders in the
heavy baryon expansions.
We see that the interference part contributes to $\gamma_2$ and $\gamma_4$,
and that their values are small in $\gamma_2$, but considerably large
in $\gamma_4$.
For the forward spin-polarizability $\gamma_0$
the contribution of the magnetic part becomes smaller by
the effect of the finite width of the $\Delta$ state,
but that of the interference part becomes large; as a result, the sum of
them
is nearly the same as that at the narrow-width limit of the $\Delta$ state.

The electric part of the $\Delta\pi$ channel is rather small in  agreement
with the result for the $\Delta\pi$ loops in HBChPT.
The magnetic  and the interference parts in the $\Delta\pi$ channel
are almost negligibly small.
It is expected that
the contributions of the $\Delta\pi$ channel are small
compared with those of the $N\pi$ channel
because of the factor $\omega^4$ in the denominator of
the dispersion relation.
We infer that the effect of the higher resonances other than
the $\Delta$ state is very small.

In Table \ref{tb:table2} we show the calculated results of
the spin-polarizabilities
with the anomalous part, and compare those with
the results of HBChPT\cite{Hemmert98} and of
the multipole analysis, where HDT and SAID refer \cite{HDT} and
\cite{Babusci98b}, respectively.
We see  good agreement with the results of HBChPT,
but the value of the $\tau_3$-independent part of $\gamma_4$ is
large compared with that of  HBChPT, and seems to
be close to the results of the multipole analysis.
This is due to the effect of the interference part between
the electric and magnetic amplitudes. As a result, the forward
spin-polarizability $\gamma_0$ is also close to those of the multipole
analysis.
Note that we cannot evaluate the proton and neutron difference, since we
did not consider the amplitude $f^{(0)}$ as it is of higher orders.
It is known that the proton and neutron difference also
cannot be predicted in HBChPT up to
the small scale expansion  of $\CO(\epsilon^3)$.

Here, let us mention about the relativistic amplitude
in Eq.~(\ref{eq:relaNpiE}). The relativistic one-loop calculation
of the $N\pi$ loops has been shown by Bernard
\etal\cite{Bernard92a}, and yields the forward spin-polarizability
$\gamma_0=2.2(3.2)\times 10^{-4}{\mathrm fm}^4$ for the proton (neutron).
The amplitude in Eq.~(\ref{eq:relaNpiE}) is found to yield
$\gamma_0=3.3\times 10^{-4}{\mathrm fm}^4$ with  dispersion
relation for relativistic invariant amplitude.
The value is
considerably  smaller than the nonrelativistic value $\gamma_0=5.1
\times 10^{-4}{\mathrm fm}^4$. Note that we must calculate
the nucleon-pole terms\cite{Drechsel92}, in order to predict
the proton-neutron difference.
The spin-polarizabilities are calculated to be
$\gamma_1=4.5,\ \gamma_2=1.6,\ \gamma_3=0.5,\ \gamma_4=-0.2$
in units of $10^{-4}{\mathrm fm}^4$, respectively.
These results are compared with the nonrelativistic values of
the electric part of the $N\pi$ channel in Table
\ref{tb:table1}, and may show the importance of relativistic approach.


In conclusion we have calculated the spin-polarizabilities
of the nucleon, where the dispersion relation
was used with the imaginary part of  Compton scattering amplitude
constructed through the unitarity condition from the pion
photoproduction amplitude in the chiral soliton model.
The pion photoproduction amplitude is given by the electric
and magnetic ones, and both the $N\pi$ and $\Delta \pi$ channels
are taken into account.
We have shown that the electric and magnetic parts in the chiral soliton
model agree with the results of the $N\pi$-loops, $\Delta$-poles
and $\Delta\pi$-loops calculated in  HBChPT.
The numerical results are also similar with each other
and qualitatively agree with those of the multipole analysis.
The interference part of the electric and magnetic amplitudes in the
$N\pi$ channel, which is not considered in HBChPT as higher orders,
is, however, large especially in $\gamma_4$,
and the resulting values of the polarizabilities are closer to
those of the multipole analysis.
The next-to-leading-order calculation with the $f^{(0)}$ amplitudes
is necessary to evaluate the difference of proton and neutron.
In this respect it is interesting to note that a large contribution
to the proton and neutron difference arises from kaon loops\cite{Butler}.

We showed that the approach with dispersion relation
is  very powerful in going further in the chiral soliton
model. A various application of such an approach is expected to follow.

\appendix
\section{Integral formulas}
\label{sec:integral}

We give angular integral formulas which are necessary to
calculate the polarizabilities:
\bey
{1\over4\pi} \int{\rm d}\Omega_q\,
{v\hat q_i \over 1-v\hat{\vect k}\!\cdot\!\hat{\vect q}}
&=&\hat k_i I_{1}(v), \nonumber \\
{1\over4\pi} \int{\rm d}\Omega_q\,
{v^2\hat q_i \hat q_j \over 1-v\hat{\vect k}\!\cdot\!\hat{\vect q}}
&=&\delta_{ij} I_{2}(v)+\hat k_i \hat k_j I_{3}(v), \nonumber \\
{1\over4\pi} \int{\rm d}\Omega_q\,
{v^3\hat q_i \hat q_j \hat q_k \over 1-v\hat{\vect k}\!\cdot\!\hat{\vect q}}
&=&\(\delta_{ij}\hat k_k+\delta_{ik}\hat k_j+\delta_{jk}\hat k_i\)I_{4}(v)
   +\hat k_i \hat k_j \hat k_k I_{5}(v) ,\nonumber \\
{1\over4\pi} \int{\rm d}\Omega_q\,
  {v^2\hat q_i \hat q_j \over
  (1-v\hat{\vect k}'\!\cdot\!\hat{\vect q})
  (1-v\hat{\vect k}\!\cdot\!\hat{\vect q})}
&=&\delta_{ij}J_{1}(v,\theta)
  +\(\hat k_i\hat k_j+\hat k_i^{\prime}\hat
k_j^{\prime}\)J_{2}(v,\theta)\nonumber \\
&&\qquad  +\(\hat k_i\hat k_j^{\prime}
+\hat k_i^{\prime}\hat k_j\)J_{3}(v,\theta), \nonumber \\
{1\over4\pi} \int{\rm d}\Omega_q\,
  {v^3\hat q_i \hat q_j \hat q_k \over
  (1-v\hat{\vect k}'\!\cdot\!\hat{\vect q})
  (1-v\hat{\vect k}\!\cdot\!\hat{\vect q})}
&=&\(\hat k_i\hat k_j \hat k_k
     +\hat k_i^{\prime}\hat k_j^{\prime} \hat k_k^{\prime}\)J_{4}(v,\theta)
  +\(\hat k_i\hat k_j\hat k_k^{\prime}
    +\hat k_i\hat k_j^{\prime}\hat k_k \right. \nonumber \\
&& \qquad \left.   +\hat k_i^{\prime}\hat k_j\hat k_k
    +\hat k_i^{\prime}\hat k_j^{\prime}\hat k_k
    +\hat k_i^{\prime}\hat k_j\hat k_k^{\prime}
    +\hat k_i\hat k_j^{\prime}\hat k_k^{\prime}\)J_{5}(v,\theta) \nonumber
\\
&+&\[\delta_{ij}(\hat k+\hat k')_k
    +\delta_{ik}(\hat k+\hat k')_j
    +\delta_{jk}(\hat k+\hat k')_i\]J_{6}(v,\theta).
\eey
The functions $I_{i}$ for $i=1,\ldots, 5$ depend on the pion velocity $v$
as follows:
\bey
I_{1}(v)&=&-1+{1\over 2v}\ln\({1+v \over 1-v}\), \nonumber \\
I_{2}(v)&=&\bhalf -{1-v^2 \over 4v}\ln\({1+v \over 1-v}\), \nonumber \\
I_{3}(v)&=&-{3\over 2}+{3-v^2 \over 4v}\ln\({1+v \over 1-v}\), \nonumber \\
I_{4}(v)&=&-{1\over 3}v^2+\bhalf
         -{1-v^2 \over 4v}\ln\({1+v \over 1-v}\), \nonumber \\
I_{5}(v)&=&{2\over 3}v^2-{5\over 2}
         +{5-3v^2 \over 4v}\ln\({1+v \over 1-v}\).
\eey
The functions $J_{i}$ for $i=1,\ldots, 6$ depend on the pion velocity $v$
and the photon scattering angle $\theta$.
We need the value for the forward scattering only, so give them at
$\theta=0$:
\footnote{ L'vov \etal\ derived
the same formulas, but their definitions are different from ours,
and theirs are also valid for $\theta\neq0$\cite{Lvov97}.}
\bey
J_{1}(v,0)&=&-1+{1\over 2v}\ln\({1+v \over 1-v}\), \nonumber \\
J_{2}(v,0)&=&{2\over 3}+{1\over 3(1-v^2)}
         -{1\over 2v}\ln\({1+v \over 1-v}\), \nonumber \\
J_{3}(v,0)&=&{1\over 3}+{1\over 6(1-v^2)}
         -{1\over 4v}\ln\({1+v \over 1-v}\), \nonumber \\
J_{4}(v,0)&=&{13\over 8}+{1\over 4(1-v^2)}
         -{15-3v^2 \over 16v}\ln\({1+v \over 1-v}\), \nonumber \\
J_{5}(v,0)&=&{13\over 24}+{1\over 12(1-v^2)}
         -{5-v^2 \over 16v}\ln\({1+v \over 1-v}\), \nonumber \\
J_{6}(v,0)&=&-{3\over 4}
         +{3-v^2 \over 8v}\ln\({1+v \over 1-v}\).
\eey

\section{Dispersion relation for $N\pi$ electric part}
\label{sec:NE}

Let $f(\omega)$ be analytic in the upper half plane of complex $\omega$
and be an odd function of $\omega$; then, we find
\ben
p\Re g(\omega)
 ={2\over\pi}\,{\rm P}\int_0^{\omega_{max}}{\rm d}\omega '\,
  {\omega ' \over \omega^{\prime 2}-\omega^2}\,\Im g(\omega ')
 +{1\over\pi}\,\Im\int_{C}{\rm d}\omega '\,
  {\omega ' \over \omega^{\prime 2}-\omega^2}\,g(\omega '), \label{eq:ds}
\een
where the path $C$ denotes the upper semi-circle path in the complex plane.
We have derived the dispersion relations in Eqs.~(\ref{eq:drpol})
at the limit of $\omega_{max}\rightarrow\infty$
and by neglecting the semi-circle integrals.
We will show that this is erroneous in the case of the electric part
for the $N\pi$ channel.

The non-Born part of the structure function of the forward scattering
amplitude, $A_5(\omega,\theta=0)$, which is equal to
$-A_6(\omega,\theta=0)$,
in HBChPT \cite{Bernard95} is given by two terms as follows:
\ben
A_{51}(\omega,0)
=-{g_A^2 m_\pi\over 8\pi^2\fpi^2} u^2
\int_0^1 dx {(1-x)^2 \sin^{-1}ux
\over \sqrt{1-u^2x^2}},
\een
and
\ben
A_{52}(\omega,0)
={g_A^2 m_\pi\over 8\pi^2\fpi^2 } u^4
\int_0^1 dx {x(1-x)^3 \over 3(1-u^2x^2)^{3/2} }
\left( \sin^{-1}ux+xu\sqrt{1-u^2x^2} \right)
\een
with $u=\omega/m_\pi$.
The integration on $x$ can be carried out and lead to
\ben
A_{51}(\omega,0) = -{g_A^2 \mpi\over 8\pi^2\fpi^2}
\left( {u^3\over 12}+{u^5\over 315}+\CO(u^7) \right),\label{eq:A51}
\een
and
\ben
A_{52}(\omega,0) = {g_A^2 \mpi\over 8\pi^2\fpi^2}
\left( {u^5\over 315}+\CO(u^7) \right),\label{eq:A52}
\een
where the $u^5$ and the higher terms are the same except for the signs
in  both functions. Therefore, the sum is given by
\ben
A_{5}(\omega,0) = A_{51}(\omega,0)+A_{52}(\omega,0) =
-{g_A^2 \mpi\over 8\pi^2\fpi^2}  {u^3\over 12},\label{eq:A5}
\een
because of complete cancellation among the higher terms.
This shows that
there is no branch cut on the real axis in the complex
$\omega$ plane; namely, there is no imaginary part of the structure function
$A_5(\omega,0)$.
However, each of $A_{51}(\omega,0)$ and $A_{52}(\omega,0)$ has
an imaginary part:
The analytic continuation to $\omega>m_\pi$ gives
\ben
\Im A_{52}(\omega,0)=-\Im A_{51}(\omega,0)
= {g_A^2 m_\pi \over 8\pi^2\fpi^2 }
{\pi\over 8u}\left(-6u\sqrt{u^2-1}+(1+2u^2)\ln{u+\sqrt{u^2-1}
\over u-\sqrt{u^2-1}}\right). \label{eq:imA52}
\een
We see that $\Im A_{52}(\omega,0)$ and $\Im A_{52}(\omega,0)$ come from
the last term in Eq.~(\ref{eq:absfEN}) by means of the identity of
Eq.~(\ref{eq:gamma7}).
Therefore, there is no contradiction with the HBChPT approach,
where the imaginary parts of
$A_5(\omega,0)$ and $A_6(\omega,0)$ are zero, if we include
the last term in Eq.~(\ref{eq:absfEN}).

Now, let us consider the dispersion relations for
$A_{51}(\omega,0)$ and $A_{52}(\omega,0)$,
\ben
\left. \Re {A_{51}(\omega,0)\over \omega^3} \right |_{\omega=0}
 ={2\over\pi}\,{\rm P}\int_0^{\omega_{max}}{\rm d}\omega '\,
  {\Im A_{51}(\omega ',0) \over \omega^{\prime 4}}
 +{1\over\pi}\,\Im\int_{C}{\rm d}\omega '\,
  {A_{51}(\omega ',0) \over \omega^{\prime 4}}, \label{eq:dsA51}
\een
\ben
\left. \Re {A_{52}(\omega,0)\over \omega^3} \right |_{\omega=0}
 ={2\over\pi}\,{\rm P}\int_0^{\omega_{max}}{\rm d}\omega '\,
  {\Im A_{52}(\omega ',0) \over \omega^{\prime 4}}
 +{1\over\pi}\,\Im\int_{C}{\rm d}\omega '\,
  {A_{52}(\omega ',0) \over \omega^{\prime 4}}. \label{eq:dsA52}
\een
From Eq.~(\ref{eq:A51}) the left-hand side (LHS) in Eq.~(\ref{eq:dsA51})
is $\displaystyle{-{g_A^2\over 8\pi^2\fpi^2}{1 \over 12\mpi^2}}$.
In the right-hand side (RHS) the dispersion integral with the imaginary
part in Eq.~(\ref{eq:imA52}) is found to be equal to the LHS
with the vanishing semi-circle integral.
On the other hand, the LHS in Eq.~(\ref{eq:dsA52}) is zero,
and the second (semi-circle) integral in the RHS cancels out the dispersion
integral in the first term.
Therefore, the dispersion relation of $A_5(\omega,0)$ can be written
as
\bey
\left. \Re {A_5(\omega,0)\over \omega^3} \right |_{\omega=0}
 &=&{1\over\pi}\,\Im\int_{C}{\rm d}\omega '\,
    {A_5(\omega ',0) \over \omega^{\prime 4}} \nonumber \\
 &=&-{g_A^2\over 8\pi^2\fpi^2}{1 \over 12\mpi^2}
    {1\over\pi}\,\Im\int_{C}{\rm d}\omega '\,{1\over \omega '} \nonumber\\
 &=&-{g_A^2\over 8\pi^2\fpi^2}{1\over 12\mpi^2},
\eey
where we used  the imaginary part of $A_5(\omega,0)$
to be zero on the real axis and
used Eq.~(\ref{eq:A5}).

The above consideration shows that the semi-circle integral cannot
be generally neglected, even though an integral on the real axis is
convergent (the integrand is zero on the real axis in the above case).
Considering a once-subtracted dispersion relation we may neglect
the integral on the semi-circle, but we cannot determine the
polarizability in this case.
This situation leads to an idea that the dispersion relation cannot
be used to obtain the polarizability,
which is a coefficient of $\omega^3$ term.
However, $A_{52}(\omega,0)$
is of $\CO(\omega^5)$, as seen in Eq.~(\ref{eq:A52}),
so that the amplitude $A_{52}(\omega,0)$
does not participate in the game, and we consider only the
dispersion relation in Eq.~(\ref{eq:dsA51}), where
the semicircle integral can be dropped.
Such an $\omega$ dependence can be inferred from the construction
as noted in section \ref{sec:sec3}.
However,  a careful treatment is necessary in such a case as
multipole analysis.


\begin{figure}
\begin{center}
\epsfig{file=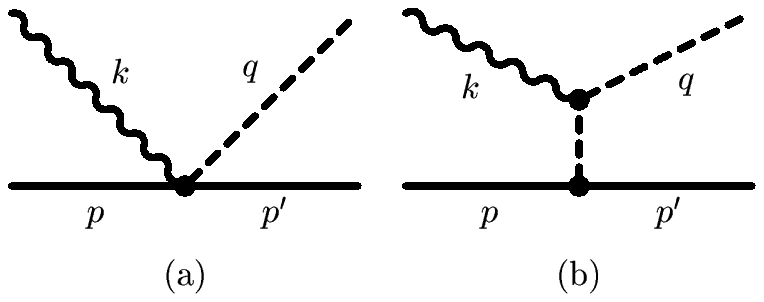}
\end{center}
\caption{Born graphs of the electric part of the
 pion photoproduction in Eq.~(\ref{eq:fEN}).
The solid, dashed and wavy lines denote the nucleon, pion and photon,
respectively. See text for the details. }
\label{fig1}
\end{figure}

\begin{figure}
\begin{center}
\epsfig{file=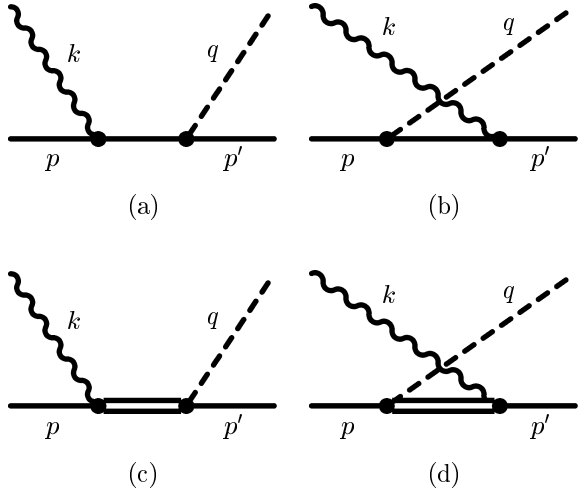}
\end{center}
\caption{Born graphs of the magnetic part of the pion photoproduction
in Eqs.~(\ref{eq:Nm}). The double solid line denote the $\Delta$ particle.
The others are the same as in Fig.~\ref{fig1}. (a) and (b) are the direct
and crossed terms of the nucleon poles, respectively. (c) and (d) are the
same as (a) and (b), respectively, but with the $\Delta$ poles.}
\label{fig2}
\end{figure}

\begin{figure}
\begin{center}
\epsfig{file=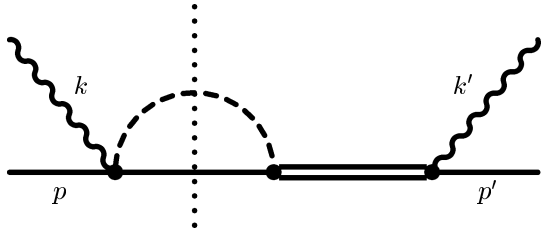}
\end{center}
\caption{
The interference between the electric amplitude
 in Fig.~\ref{fig1}(a) and  the magnetic one in Fig.~\ref{fig2}(c).
The vertical dotted line denotes the on-shell.}
\label{fig3}
\end{figure}

\begin{figure}
\begin{center}
\epsfig{file=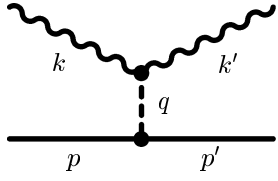}
\end{center}
\caption{The anomalous $\pi_0\gamma\gamma$ term.}
\label{fig4}
\end{figure}

\begin{table}
\caption{
Calculated spin-polarizabilities of the nucleon in the chiral soliton model
without the anomalous term due to the $\pi_0\gamma\gamma$ process.
For a comparison those in HBChPT are also shown.
Parameters in the chiral soliton model
are taken to be empirical ones, except for the $N\Delta$ transition
parameters predicted in the soliton model.
For the chiral soliton model
$E$, $M$ and $I$ denote  the contributions of the electric, magnetic
and interference parts, respectively.
In the parentheses  the values at the narrow-width limit of the
$\Delta$ particle are given in the chiral soliton model.
For the results of HBChPT
the numbers in the parentheses are the values with the old estimation
of $g_{\pi N\Delta}$ and $b_1$.
All values are in unit of $10^{-4} \mathrm{fm}^4$.
}
\begin{center}
\begin{tabular}{dddddddddd}
&\multicolumn{6}{c}{Chiral Soliton model} &
\multicolumn{3}{c}{HBChPT\cite{Hemmert98}}
\\ \hline
&\multicolumn{3}{c}{$N\pi$ channel} &
\multicolumn{3}{c}{$\Delta\pi$ channel} & & &  \\
& $E$ & $M$ & $I$ & $E$ & $M$ & $I$
& $N\pi$-loop & $\Delta$-pole & $\Delta\pi$-loop  \\ \hline
$\gamma_1$ & 5.1    & 0.0      & 0.0
           & $-$0.4 & 0.0      & 0.0
           & 4.56   &0         & $-$0.21
               \\
           &        &          &
           &        &          &
           &        &          & ($-$0.4)
                   \\
$\gamma_2$ & 2.5    & $-$2.5   & $-$0.1
           & $-$0.4 & 0.1      & 0.5
           & 2.28   & $-$2.40  & $-$0.23
               \\
           &        & ($-$4.0) & ($-$0.3)
           &        &          &
           &        & ($-$4.0) & ($-$0.5)
                   \\
$\gamma_3$ & 1.3    & 0.0      & 0.0
           & $-$0.2 & 0.0      & 0.0
           & 1.14   & 0        & $-$0.12
               \\
           &        &          &
           &        &          &
           &        &          & ($-$0.2)
                   \\
$\gamma_4$ & $-$1.3 & 2.5      & 1.2
           & 0.2    & $-$0.1   & $-$0.1
           & $-$1.14& 2.40     & 0.12
            \\
           &        & (4.0)    & (0.5)
           &        &          &
           &        & (4.0)    & (0.2)
                   \\ \hline
$\gamma_0$ & 5.1    & $-$2.5   & $-$2.4
           & $-$0.4 & 0.1      & $-$0.3
           & 4.5    & $-$2.4   & $-$0.2
               \\
           &        & ($-$4.0) & ($-$0.8)
           &        &          &
           &        & ($-$4.0) & ($-$0.4)
                   \\ 
\end{tabular}
\end{center}
\label{tb:table1}
\end{table}

\begin{table}
\caption{ Spin-polarizabilities of the nucleon in the chiral soliton model
with the anomalous term.
The results in HBChPT and of the multipole analysis
are also shown.  All values are in unit of $10^{-4} \mathrm{fm}^4$.}
\begin{center}
\begin{tabular}{dr@{}lr@{}lddddd}
& & & & & \multicolumn{5}{c}{multipole analysis} \\
& \multicolumn{2}{c}{Chiral Soliton} &
\multicolumn{2}{c}{HBChPT\cite{Hemmert98}}
& \multicolumn{2}{c}{HDT\cite{HDT}} &
\multicolumn{2}{c}{SAID\cite{Babusci98b}}
& \multicolumn{1}{c}{$\pi^0$ exch.} \\
& & & & & p & n & p & n & \\ \hline
$\gamma_1$ & 4.7    & $-$22.8$\tau_3$ & 4.4    & $-$21.7$\tau_3$
           & 5.1    & 6.1    & 3.1    & 6.3    & $-$22.5$\tau_3$ \\
$\gamma_2$ & 0.1    &                 & $-$0.3 &
           & $-$1.1 & $-$0.8 & $-$0.8 & $-$0.9 & \\
$\gamma_3$ & 1.1    &   +11.4$\tau_3$ & 1.1    & +10.9$\tau_3$
           & $-$0.6 & $-$0.6   & 0.3  & $-$0.7 &    11.2$\tau_3$ \\
$\gamma_4$ & 2.5    & $-$11.4$\tau_3$ & 1.3    & $-$10.9$\tau_3$
           & 3.4    & 3.4    & 2.7    & 3.8    & $-$11.2$\tau_3$ \\ \hline
$\gamma_0$ & $-$0.1 &        & 2.0    &
           & $-$0.6 & $-$0.2 & $-$1.5 & $-$0.4 & \\ 
\end{tabular}
\end{center}
\label{tb:table2}
\end{table}



\begin{references}


\bibitem{DHG}
S.D. Drell and A.C. Hearn,
Phys. Rev. Lett. {\bf 16} (1966) 908;
S. Gerasimov,
Sov. J. Nucl. Phys. {\bf 2} (1966) 430.

\bibitem{GGT}
M. Gell-Mann, M.L. Goldberger and W.E. Thirring,
Phys. Rev. {\bf 95} (1954) 1612.

\bibitem{Ragusa}
S. Ragusa,
Phys. Rev. {\bf D47} (1993) 3757;
Phys. Rev. {\bf D49} (1994) 3157.

\bibitem{Sandorfi94}
A.M. Sandorfi, C.S. Whisnant and M. Khandaker,
Phys. Rev. {\bf D50} (1994) R6681.

\bibitem{Bernard92a}
V. Bernard, N. Kaiser, J. Kambor and Ulf-G. Meissner,
Nucl. Phys. {\bf B388} (1992) 315.


\bibitem{Hemmert97}
T.R. Hemmert, B.R. Holstein and J. Kambor,
Phys. Rev. {\bf D55} (1997) 5598.

\bibitem{Hemmert98}
T.R. Hemmert, B.R. Holstein, J. Kambor and G. Kn\"ochlein,
Phys. Rev. {\bf D57} (1998) 5746.

\bibitem{Tanushi}
Y. Tanushi, Y. Nakahara, S. Saito and M. Uehara,
Phys. Lett. {\bf B425} (1998) 6.

\bibitem{Saito95}
S. Saito and M. Uehara,
Phys. Rev. {\bf D51} (1995) 6059.

\bibitem{Adkins}
G.S. Adkins, C.R. Nappi and E. Witten,
Nucl. Phys. {\bf B288} (1983) 552;
G.S. Adkins and C.R. Nappi,
Nucl. Phys. {\bf B233} (1984) 109.

\bibitem{Saito93}
S. Saito, F. Takeuchi and M. Uehara,
Nucl. Phys. {\bf A556} (1993) 317.

\bibitem{Bernard92b}
V. Bernard, N. Kaiser and Ulf-G. Meissner,
Nucl. Phys. {\bf B373} (1992) 346.

\bibitem{Kniehl}
B.A. Kniehl,
Acta Phys. Pol. {\bf B27} (1996) 3631.

\bibitem{Wess}
J. Wess and B. Zumino,
Phys. Lett. {\bf 37B} (1971) 95;
E. Witten,
Nucl. Phys. {\bf B223} (1983) 422.

\bibitem{Kaiser92}
V. Bernard, N. Kaiser and Ulf-G. Meissner,
Nucl. Phys. {\bf B383} (1992) 442;
V. Bernard, N. Kaiser, T.S.H. Lee and Ulf-G. Meissner,
Phys. Rep. {\bf 246} (1994) 316.

\bibitem{Drechsel92}
D. Drechsel and L. Tiator,
J. Phys. G : Nucl. Part. Phys. {\bf 18} (1992) 449.

\bibitem{Dashen94}
R. Dashen, E. Jenkins and A.V. Manohar,
Phys. Rev. {\bf D49} (1994) 4714;
E. Jenkins and A.V. Manohar,
Phys. Lett. {\bf B335} (1994) 452.

\bibitem{Lvov98}
A.I. L'vov,
nucl-th/9810032

\bibitem{HDT}
D. Drechsel, G. Krein and O. Hanstein,
Phys. Lett. {\bf 420B} (1998) 248.

\bibitem{Babusci98b}
D. Babusci, G.Giordano, A.I. L'vov, G.Matone and A.M. Nathan,
Phys. Rev. {\bf C58} (1998) 1013.

\bibitem{Butler}
M.N. Butler and M.J. Savage,
Nucl. Phys. {\bf B399} (1993) 69.

\bibitem{Lvov97}
A.I. L'vov, V.A. Petrun'kin and M. Schumacher,
Phys. Rev. C {\bf C55} (1997) 359

\bibitem{Bernard95}
V.Bernard, N.Kaiser and Ulf-G. Meissner,
Int. J. Mod. Phys. {\bf E4} (1995) 193.

\end{references}
\end{document}